\documentclass[aps,prl,twocolumn,showpacs]{revtex4}
\usepackage{amssymb}

\usepackage{color}
\usepackage{graphicx}
\usepackage{cancel}
\usepackage{bm}
\usepackage{dcolumn}

\def \tr      {\rm{tr}}

\def\rv{{\bf r}}

\def\sv{{\bf s}}
\def\kv{{\bf k}}
\def\nv{{\bf n}}
\def\pv{{\bf p}}
\def\qv{{\bf q}}

\def\Bv{{\bf B}}

\def\Sv{{\bf S}}

\def\Acal{{\mathcal A}}
\def\Ucal{{\mathcal U}}
\def\Im{{\rm Im}}

\begin{document}

\title{Spin dynamics of cold fermions with synthetic spin-orbit coupling}
\author{I. V. Tokatly}

\affiliation{Nano-bio Spectroscopy group and ETSF Scientific
Development Centre, Departamento de F\'isica de Materiales, Universidad del Pa\'is
Vasco UPV/EHU, E-20018 San Sebast\'ian, Spain} 
\affiliation{IKERBASQUE Basque Foundation for Science, 48011, Bilbao, Spain}

\author{E. Ya. Sherman}
\affiliation{Department of Physical Chemistry, Universidad del Pa\'is Vasco UPV/EHU,
48080 Bilbao, Spain}
\affiliation{IKERBASQUE Basque Foundation for Science, 48011, Bilbao, Spain}

\begin{abstract}
We consider spin relaxation dynamics in cold Fermi gases with a pure-gauge spin-orbit
coupling corresponding to recent experiments. We show that such experiments can give
a direct access to the collisional spin drag rate, and establish conditions for the observation 
of spin drag effects. In the recent experiments the dynamics is found to be mainly ballistic leading to 
new regimes of reversible spin relaxation-like processes. 

\end{abstract}

\maketitle

The development of spintronics, the branch of physics studying
spin-determined dynamical and transport phenomena was mainly related to solid-state structures. In these systems,
spin-orbit coupling (SOC), one of the key elements of spintronics, is 
well-understood \cite{Winkler03,Zutic04,Dyakonov08,Wu10} and many
interesting spin-related effects have been studied experimentally and
theoretically. Very recently { a detailed study of spin dynamics in ultracold atomic gases has 
become experimentally feasible \cite{Sommer,Lin,Wang,Cheuk}. 
In particular, two systems where SOC is produced by a special
design of optical fields, attracted a great deal of attention.} One of them
is the spin-orbit coupled Bose-Einstein condensates \cite{Stanescu,Lin},
with the pseudospin $1/2$ degree of freedom. The other class is represented by the cold fermion
isotopes $^{40}$K in Ref.~\cite{Wang} and much lighter $^{6}$Li studied in
Ref.~\cite{Cheuk}. In both cases, in addition to the SOC, an
effective Zeeman magnetic field can be produced optically.

Typically in solids, e.g. in doped semiconductors, a disorder, randomizing motion of electrons, plays the dominant role in the spin dynamics. 
The electron-electron collisions become crucial only at high temperatures, or in intrinsic 
semiconductors with optically pumped electrons and holes \cite{eeinter}. From
this point of view, cold atomic gases offer a unique possibility of
seeing basic effects of interactions in the pure form since the disorder is
absent there. The interatomic collisions lead to the spin drag determining the spin diffusion and, as we
will see below, can be important for the spin dynamics in cold Fermi
gases with SOC.

It is well-appreciated that in the presence of strong SOC the effects of interatomic
interactions in the spin dynamics are difficult to analyze as this requires
tracing essentially coupled orbital and spin subsystems. 
Fortunately, these dynamics become uncoupled not only for vanishing SOC, 
but also when it corresponds to an effective non-Abelian vector potential \cite
{Mineev92,Frolich93,Aleiner01,Levitov03,Lyanda,Yang06,Liu07,Natano07,Yang08,Leurs08,Raimondi,Tokatly08,Tokatly10}
of a pure gauge form, which happens in a broad class of systems. Remarkably, the three-dimensional (3D) 
fermionic gases with SOC realized in recent experiments \cite{Wang,Cheuk,Liu} belong to this interesting class. 
For a pure gauge SOC the behavior of the physical system maps to that of a system without
SOC, which allows to consider effects of SOC of an
arbitrary strength. In this case all qualitative features of the spin
dynamics are the same as for a generic SO field, but the analysis is
much easier. Here we study spin dynamics for systems
with a pure gauge SO coupling, where the entire pattern even if it is
complicated by the interatomic interactions, can be explored using a single
formula. As a byproduct of our analysis we show that studying the spin relaxation in SO coupled 
systems may offer an alternative route to experimentally access the spin drag resistivity in SOC-free systems.

We represent the Hamiltonian of a 3D Fermi gas of atoms with the mass $m$ and a linear in momentum SOC
as follows \cite{Tokatly08,Tokatly10}: 
\begin{equation}
 \hat{H} = \int d^{3}r
\left\{
{\bm\psi}^{\dagger}
\left[
\frac{\left(i\nabla_j + \Acal_j^a\sigma^a\right)^2}{2m}+B^{b}\sigma^{b}
\right]
{\bm\psi}
\right\}
+ \hat{H}_{\rm int},
 \label{H}
\end{equation}
where ${\bm\psi}$ is a spinor fermionic field operator, $\sigma^{a}$ with $a=\{x,y,z\}$ are the Pauli matrices, 
the space components $\Acal_{j}^{a}$ of an effective SU(2) potential parametrize SOC, and $B^{b}$ 
is the Zeeman field (the time component of the SU(2) potential) \cite{note1}. 
The term $\hat{H}_{\rm int}[{\bm\psi}^{\dagger},{\bm\psi}]$ in Eq.~(\ref{H}) describes 
the interparticle interactions and the external trapping potential. In the following 
we assume a short-range interaction characterized by a $s$-wave scattering length $a_{s}$. 
For simplicity we consider a homogeneous system, assuming the validity 
of the local approximation.

The spin relaxation is probed by applying the Zeeman field of the form $\Bv(t)=\Bv\theta(-t)$. 
That is, we first bring the system to an equilibrium polarized state in the presence of the 
static Zeeman field which is then switched off at $t=0$. After the polarizing field is 
released, the spin polarization relaxes to zero because of the combined action of the 
SOC-induced spin precession and the particle-particle collisions. 

In conventional semiconductors the SOC is usually weak, that
is $\left|\Acal_j^a\right|\ll p_{\rm F},$ where $p_{\rm F}$ is the Fermi momentum. As a result, only { the
vicinity of the Fermi surface is important for} the spin dynamics. Moreover, the weak coupling practically 
does not influence the initial spin-polarized state. With optical fields one can achieve the 
ultrastrong coupling where the effective 
SU(2) field $\left|\Acal_{j}^{a}\right|$ is close to or even larger than the Fermi- 
or the mean thermal momentum. In this ultrastrong coupling regime the collisional relaxation
processes can become irrelevant and the dynamics is mainly due to the
nonuniform spin precession, that is its broadening in the
momentum space. In addition, the strong SOC significantly modifies the initial state making it 
strongly non-semiclassical.

While the description of the spin dynamics for a strong generic SOC is very complicated, significant 
simplifications occur in the special case of a pure gauge SO field of the following ``direct product'' form
\begin{equation}
\label{pure-gauge}
\Acal_{j}^{a} = \frac{1}{2}q_{j}n^{a}.
\end{equation}
Here $\nv$ is a unit vector pointing along the SO precession axis 
(momentum independent in this case), and the vector $\qv$ determines the strength of SOC
for every direction of the particles' momentum coupled to the spin degrees of freedom. Exactly 
this type of the SU(2) field has been realized in the recent 
experiments with cold Bose and Fermi gases \cite{Lin,Wang,Cheuk}.

Because different components of the SU(2) potential in Eq.~(\ref{pure-gauge}) commute, 
it can be gauged away by a local SU(2) rotation of the fermion fields:
\begin{equation}
 {\bm\psi}(\rv) = \Ucal(\rv)\widetilde{{\bm\psi}}(\rv), \quad \Ucal(\rv) = \exp\Big[\frac{i}{2}q_{j}r_{j}n^{a}\sigma^{a}\Big].
 \label{rotation}
\end{equation}
{ After the rotation,} the Hamiltonian of Eq.~(\ref{H}) reads
\begin{equation}
\hat{H} = \int d^3{r}
\left\{
\widetilde{\bm\psi}^{\dagger}
\left[
-\frac{\nabla^2}{2m}+{\bm\sigma}\widetilde\Bv(\rv)
\right]
\widetilde{\bm\psi}
\right\} 
+ \hat{H}_{\rm int}[\widetilde{\bm\psi}^{\dagger},\widetilde{\bm\psi}],
 \label{tilde-H}
\end{equation}
where $\widetilde\Bv(\rv)=\tr\{{\bm\sigma}\Ucal^{-1}(\rv)({\bm\sigma}\Bv)\Ucal(\rv)\}/2$ 
is the transformed Zeeman field. For definiteness we consider 
the configuration of fields generated in the experiments of Ref.~\cite{Wang}. 
This corresponds to $\nv=(0,0,1)$, $\qv=(q,0,0)$, and $\Bv=(B,0,0)$, which translates 
to the following helicoidal structure \cite{Bernevig06,Liu06,Koralek09,Slipko}  of the transformed Zeeman field 
\begin{equation}
 \label{tilde-B}
 \widetilde\Bv(\rv) = B\left[ 
\widehat{\mathbf{x}}\cos (qx)+\widehat{\mathbf{y}}\sin (qx)\right] .
\end{equation}
Since the interparticle interaction and the trap potential are spin independent 
the form of $\hat{H}_{\rm int}$ is untouched by the SU(2) rotation of Eq.~(\ref{rotation}). 
Therefore the physical system with the pure gauge SOC subjected to a uniform 
polarizing field $\Bv=\widehat{\mathbf{x}}B$ is mapped to a dual system without SOC
in the presence of an inhomogeneous Zeeman field -- the magnetic helix of Eq.~(\ref{tilde-B}). 

The mapping of the spin relaxation for a pure gauge SOC to the diffusive 
spin dynamics in the SO-free system has been already noted in the context of conventional
semiconductors \cite{Tokatly08,Tokatly10} where the spin relaxation/diffusion 
is mostly determined by a disorder. Apparently this is an absolutely general statement, 
which in particular implies that the spin diffusion coefficient $D_{\rm s}$ and the spin 
drag rate $\Gamma_{\rm s}$ in cold atomic gases can be measured by monitoring the 
dynamics of a global averaged spin, provided the proper SOC in created. 

After the transformation of Eq.~(\ref{rotation}) the spin dynamics can be 
easily found using the linear response theory for homogeneous, SOC-free systems. 
Namely, we consider a response to a week Zeeman field Eq.~(\ref{tilde-B}) with $B(t)=B\theta(-t)$. 
The spin density $\widetilde\Sv(\rv,t)$ in the transformed system follows the helicoidal 
pattern of Eq.~(\ref{tilde-B}) with a time-dependent amplitude $S(t)$: 
\begin{equation}
 \label{helix}
  \widetilde\Sv(\rv,t) = S(t)\left[ 
\widehat{\mathbf{x}}\cos (qx)+\widehat{\mathbf{y}}\sin (qx)\right] .
\end{equation}
The amplitude $S(t)$ of the spin helix can be related to the standard 
spin response function $\chi_{\sigma\sigma}(q, \omega)$ at 
the wave vector $\qv$ (see, e.g., Ref.~\cite{Forster}) 
\begin{equation}
 \label{S-general}
 S(t) = S(0)\int_{-\infty}^{\infty}\frac{d\omega}{2\pi i}
 \left[ \frac{\chi_{\sigma \sigma }(q,\omega )}{\chi _{\sigma \sigma }(q,0)}-1\right]
 \frac{e^{-i\omega t}}{\omega }.
\end{equation}
To recover the spin dynamics in the original system we apply the inverse SU(2) 
transformation to the spin distribution $\widetilde\Sv(\rv,t)$ in Eq.~(\ref{helix}). 
This unwinds the helix and leads to the evolutions of the spatially 
uniform physical spin distribution $\Sv(t) = S(t)\widehat{\mathbf{x}}$ presented below.

Equation (\ref{S-general}) relates spin dynamics 
in the physical system with SOC to the smearing out of an inhomogeneous spin texture in a system without it. 
We emphasize that Eq.~(\ref{S-general}) is exact and valid for 
any phase without spontaneous ferromagnetic ordering, either normal or superfluid one.
The SOC enters the spin dynamics solely via the wave vector dependence of the spin-spin 
correlation function $\chi_{\sigma\sigma}(q,\omega)$ of the SOC-free system. 
Hence by monitoring the spin evolution for different values of $q$ one 
can access the entire momentum and frequency dependence of $\chi_{\sigma\sigma}(q,\omega)$. 

We begin with an illustration of the simplest universal type of 
spin dynamics occurring for small SO couplings, $v_{\rm at}q\ll\Gamma_{\rm s}$, where
$v_{\rm at}$ is the typical atomic velocity such as the Fermi velocity $v_{\rm F}$ or the mean 
thermal velocity $v_{T}$ in the degenerate and nondegenerate system, respectively. 
In this case the expression in the square brackets in Eq.~(\ref{S-general}) 
takes the diffusive form $i\omega/(i\omega + D_{\rm s}q^2)$,
which is guaranteed by the spin conservation law. Hence the spin relaxes exponentially $S(t)=S(0)\exp{\left(-D_{\rm s}q^2t\right)}$
by the Dyakonov-Perel' mechanism \cite{Dyakonov73}. 
The relaxation rate for a given SOC is directly determined by the spin 
diffusion coefficient of the dual SOC-free system.

For an arbitrary strong coupling the dynamics is much richer. To explore the behavior in 
the full range of SOC, densities, and temperatures, we employ a spin version of 
the ``conserving'' relaxation time scheme by Mermin \cite{Mermin}. In this scheme 
one approximates the collision integral in the 
linearized kinetic equation for the spin part $\sv_\pv(\qv,t)$ of the Wigner function by the following form,
 ${\bf I}_{\rm col} = -\Gamma_{\rm s}[\sv_\pv(\qv,t) - \sv_\pv^{\rm eq}(\qv,t)]$,
where $\sv_\pv^{\rm eq}(\qv,t)$ is the equilibrium function corresponding to the instantaneous 
value of the average spin $\Sv(\qv,t)$. For any inhomogeneity wave vector $\qv$ this 
approximation recovers the correct static and high-frequency response, and respects the 
local spin conservation. The dynamical spin response function in this scheme takes the form
\begin{equation}
\chi _{\sigma \sigma }(q,\omega)=
\frac{\chi _0(q,\omega+i\Gamma_{\rm s})}
{1-\displaystyle{\frac{i\Gamma_{\rm s}}{\omega+i\Gamma_{\rm s}}}
\left[ 1-\displaystyle{\frac{\chi_0(q,\omega+i\Gamma_{\rm s})}
{\chi_0(q,0)}}\right]},
\end{equation}
where $\chi_0(q,\widetilde\omega)$ is the Lindhard function
\begin{equation}
\label{chi0}
\chi _0(q,\widetilde{\omega})=\sum_{\pv}
\frac{f_{\pv+\qv/2} - f_{\pv-\qv/2}}{\widetilde{\omega} - (\epsilon_{\pv+\qv/2}-\epsilon_{\pv-\qv/2})}
\end{equation} with $f_{\pv}$ being the Fermi function, and $\epsilon_{\pv}=p^{2}/2m$.

The spin drag relaxation rate $\Gamma_{\rm s}$ enters the Mermin's 
scheme as a phenomenological parameter originating 
from interatomic collisions. Although collisions conserve 
the total momentum, they cause a friction between different spin species, and thus lead to relaxation
and damped response in the spin channel. Strictly speaking, the spin drag rate $\Gamma_{\rm s}$ depends on $q$. 
It is however clear that for large $q\gtrsim p_{\rm F}$ the average spin relaxes mostly 
because of fast nonuniform precession (ballistic motion). The spin drag contribution 
becomes qualitatively important in the opposite limit of $q\ll p_{\rm F}$. Therefore below we adopt 
the standard expression based on the Born approximation at $q=0$ \cite{Amico,Polini07}
\begin{equation}
\label{drag}
\Gamma_{\rm s}=\left(\frac{4\pi a_s}{m}\right)^2
\sum_{\kv}\frac{k^{2}}{3mnT}\int_{0}^{\infty }\frac{d\omega}{\pi}
\frac{\left[\Im\chi_{0}(k,\omega)\right]^{2}}{\sinh ^{2}\left(\omega/2T\right) },  
\end{equation}
where $T$ is the temperature, and $n$ is the concentration of the atoms.
The spin drag rate of Eq.~(\ref{drag}) has the following characteristic temperature dependence
\begin{eqnarray}\label{Gamma-low-T}
\Gamma_{\rm s}\sim E_{\rm F}\left( p_{\rm F}a_{s}\right)^{2} \times
\left\{
\begin{array}{l}
\left(T/E_{\rm F}\right)^{2},\qquad T\ll E_{\rm F} \\
\\
\left(T/E_{\rm F}\right)^{1/2},\qquad T\gg E_{\rm F}.
\end{array}
\right.
\label{Gamma_s_T}
\end{eqnarray}
The low-$T$ behavior is typical for the Fermi liquid, while in the
high-$T$ limit it describes the frequency of collisions between particles with a scattering 
cross-section $\sim a_{s}^{2}$ moving at the mean thermal velocity $v_{T}\sim\sqrt{T/m}$. 
For the repulsive interaction the Stoner instability criterion limits
the product $p_{\rm F}a_{s}<\pi /2$ \cite{Duine}, thus preventing one from having a
fast relaxation in the degenerate gas.
     
\begin{figure}[h]
\includegraphics*[width=6cm]{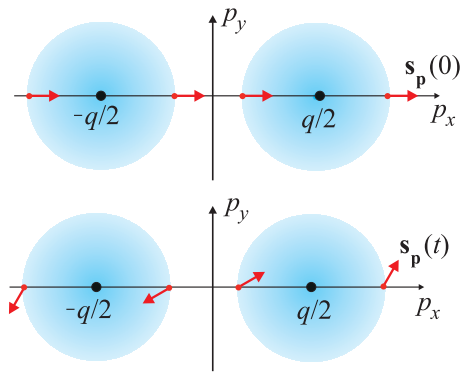}
\caption{(Color online.) Schematic plot of 
precession of spins of atoms with different momenta ${\mathbf s}_{\mathbf p}(t)$, as shown by
short arrows, in the case of very strong SOC {with $q>2p_F$}. Filled circles show spin-split thermally broadened 
Fermi spheres.}
\label{split_FS}
\end{figure}

Equations (\ref{S-general})-(\ref{drag}) provide us with a simple unified description of the spin dynamics 
in a wide range of temperatures, SO couplings, densities and interactions. 
This scheme captures all main physical effects which influence the spin relaxation. 
The difference in the Fermi functions in Eq.~(\ref{chi0}) describes the modification of 
the initial state by the SO coupling. It selects a part of the momentum space with the particles 
polarized by the initial Zeeman field, as it is shown in Fig.~\ref{split_FS} for the very strong coupling. 
The difference in energies in the denominator in Eq.~(\ref{chi0}) 
determines the frequency of the spin precession, while the imaginary shift of $\omega$ by $i\Gamma_{\rm s}$ 
describes the spin drag effect. 
% The case of strong SOC is shown in Fig. \ref{split_FS}.  
Now we use Eqs.~(\ref{S-general})-(\ref{drag}) to explore different regimes of spin dynamics and
discuss those not achievable in the solid-state experiments. 

We consider first the low-temperature degenerate gas, $T\ll E_{\rm F}$. If SOC is weak, 
so that $q\ll p_{\rm F}$, only particles near the Fermi surface participate in the dynamics. 
The characteristic time scale is set up by $1/qv_{\rm F}$, that is the time for ballistically 
traversing the period of the spin helix. The relation of this time to the spin drag rate $\Gamma_s$ determines 
the regime of the spin dynamics. If $qv_{\rm F}/\Gamma_{\rm s}\gg 1$, the dynamics is ballistic and the 
evolution of the total spin is dominated by the inhomogeneous $\mathbf{p}$-dependent spin precession 
of the Fermi-surface particles. Apparently this is always the case at sufficiently low temperatures, $T\to 0$. 
In the opposite limit of $qv_{\rm F}/\Gamma_{\rm s}\ll 1$ the dynamics is diffusive, which corresponds to 
the conventional Dyakonov-Perel' spin relaxation mechanism \cite{Dyakonov73}. From Eq.~(\ref{Gamma-low-T}) we find that with the
increase of the temperature the ballistic regime crosses over to the diffusive one at 
$T_{\rm db}\sim (v_{\rm F}/a_{s})\sqrt{q/p_{\rm F}}$. Condition $T_{\rm db}\ll E_{\rm F}$ implies
$q\ll p_{\rm F}\left(p_{\rm F}a_{s}\right)^{2}$ for the SOC strength. Taking into account that, e.g., in
Ref.~\cite{Wang}, { $p_{\rm F}a_{s}\sim 10^{-2}$, we conclude that to reach the diffusive 
regime one has to increase the interaction or to make the SOC considerably weaker.}

\begin{figure}[h]
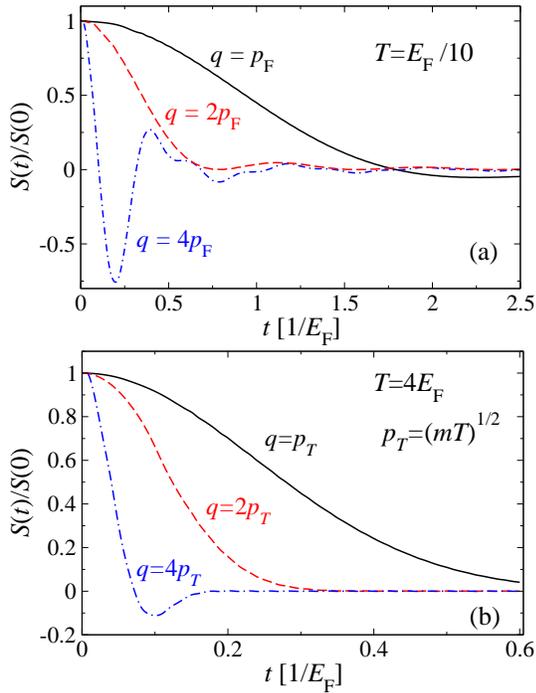

\includegraphics*[width=7cm]{large_Q_low_T.eps}\\
\includegraphics*[width=7cm]{large_Q_high_T.eps}
\caption{(Color online.)  Time dependence of the total spin at strong spin-orbit coupling. 
(a) low temperature, degenerate gas, (b) high temperature, classical gas. The values of $q$ and $T$  
are shown in the plots.}
\label{large_q}
\end{figure}

Apart from determining the spin drag rate, the temperature effects modify 
the dynamics via broadening the Fermi distribution in Eq.~(\ref{chi0}). 
Indeed, the broadening of the velocity distribution by $\delta
v_{\rm F}\sim \left( T/E_{\rm F}\right) v_{\rm F}$ causes the broadening in the
precession rate by the typical value of  $\delta\Omega=\left(T/E_{\rm F}\right) qv_{\rm F}$. 
In the ballistic regime the resulting time dependence of the total spin is
\begin{equation}
S\left(t\right) =S(0)\frac{\sin (v_{\rm F}qt)}{\sinh (\pi t\delta \Omega )}%
\frac{\pi T}{E_{\rm F}},
\end{equation}
where the rapid increase in $\sinh (\pi t\delta \Omega )$ at $t>1/\delta \Omega $ leads to the exponential
damping \cite{Glazov07} in the spin oscillations initially following the $\sin (v_{\rm F}qt)/v_{\rm F}qt$ time
dependence. At $\delta \Omega/\Gamma_{\rm s} \lesssim 1,$ that is at 
$T\gtrsim T_{\rm db}^{2}/E_{\rm F}$ the asymptotic behavior is
determined by $\Gamma_{\rm s}$ rather than by $\delta \Omega$.

For strong SOC the dynamics is mostly ballistic, but still quite nontrivial and diverse. 
As it can bee seen in Fig.\ref{large_q}(a), at $q=p_{\rm F}$ 
the spin relaxes to zero monotonically, while pronounced large amplitude 
oscillations appear for ultrastrong SOC with $q\gg p_{\rm F}$. 
The reason is that in the latter case we have two well separated 
spin-split Fermi-spheres. The position of the sphere center 
determines the mean precession rate, and the
broadening is determined by the Fermi momentum $p_{\rm F}$.
An interesting intermediate regime occurs at $q=2p_{\rm F}$ when the spin-split
Fermi-surfaces touch each other and the time dependence of the spin takes the form 
\begin{equation}
S\left(t\right) =S(0)\frac{\sin^{2}(4E_{\rm F}t)}{(4E_{\rm F}t)^{2}}.
\end{equation}

Now we consider the high-temperature case, $T\gg E_{\rm F}$. In this limit the mean free path $\sim v_T/\Gamma_s$
does not depend on the temperature [see Eq.~(\ref{Gamma_s_T})]. Therefore the dynamical regime is
temperature-independent -- the dynamics is
ballistic if $(q/p_{\rm F})/\left( p_{\rm F}a_{s}\right) ^{2}\gg 1$ and
diffusive otherwise. The condition of diffusive relaxation is 
equivalent to the condition $T_{\rm db}\ll E_{\rm F}$.  
As a result, if $q\ge p_{\rm F}\left( p_{\rm F}a_{s}\right)^{2}$, the spin dynamics is 
ballistic at any temperature. In this case the width of the thermal 
Maxwell distribution is much larger than the scale of SOC and the 
inhomogeneous precession is equivalent to the Fourier
transform of the Gaussian momentum distribution with the coordinate $qt/m$.
For a weak SOC in the ballistic regime the spin relaxation is
Gaussian $S(t)=S(0)\exp\left[-\left(v_{T}qt\right)^{2}/6\right]$ 
with the time scale of $1/qv_{T}$.  At larger scattering length $a_{s}$, that is at 
larger $\Gamma_{\rm s}$, we enter the diffusive regime where the Gaussian damping slows down to the
exponential Dyakonov-Perel' relaxation with the timescale $\Gamma_{\rm s}/q^{2}v_{T}^{2}$. 
In the nondegenerate gas the function $S(t)$ is practically always monotonic. 
To detect signatures of the oscillatory spin precession one needs to go to the ultrastrong SOC and to satisfy 
experimentally the condition $q\gg p_{T}$.

It is worth noting that in general for the weak SOC the spin dynamics in the ballistic limit
is determined by the parameter $v_{\rm at}q$ 
since this is the only characteristics of the low-energy excitation spectrum entering  Eq.~(\ref{chi0}). 
For the strong coupling with clearly 
separated spin-split momentum distributions we have two main 
parameters describing spin dynamics: spin precession rate 
of the order of $q^{2}/2m$ and spin relaxation rate of the order of 
$qv_{\rm at}$, as can be seen from comparison of Fig.\ref{large_q}(a) and (b) for
the low- and high-temperature gases, respectively. This links the spin relaxation 
dynamics to the spectrum of electron-hole excitations in the Fermi gas.

We emphasize that the above limiting regimes of the spin dynamics are described within a single unified scheme, 
which is completely determined by { four} experimentally controllable parameters, the density $n$, the temperature $T$, 
the scattering length $a_s$, { and the SOC strength $q$}. Apparently all crossovers and transition regimes are also covered by this approach.

To conclude, we have shown that the spin dynamics in cold Fermi gases can be
mapped on the spectrum of { particle-hole excitations. The spin drag rate
can be obtained from the relaxation of the uniform spin density without
generating an inhomogeneous spin distribution in the real-space, provided the conditions of diffusive spin dynamics in terms of the SOC strength, concentration, and the scattering length are satisfied.}

IVT acknowledges funding by the Spanish MEC (FIS2007-65702-C02-01) and ''Grupos
Consolidados UPV/EHU del Gobierno Vasco'' (IT-319-07). This work of
EYS was supported by the University of Basque Country UPV/EHU under program UFI 11/55,
Spanish MEC (FIS2012-36673-C03-01), and ''Grupos Consolidados
UPV/EHU del Gobierno Vasco'' (IT-472-10).


\begin{thebibliography}{99}
\bibitem{note1} The units with $\hbar\equiv 1$  are employed and the summation over repeated indices 
is assumed throughout the paper.
\end{thebibliography}
\end{document}